\documentclass[aps,prc,10pt,showpacs,showkeys,twocolumn,superscriptaddress,groupedaddress]{revtex4-2}
\usepackage[latin1]{inputenc}
\usepackage[english]{babel}
\usepackage{graphicx,psfrag}
\usepackage{amsmath}
\usepackage{amssymb}
\usepackage{amscd}
\usepackage{eucal}
\usepackage{color}
\usepackage{bm}
\usepackage{braket}
\usepackage{dsfont}
\begin{document}
\title{Translationally invariant matrix elements of general one-body operators}

\author{Petr Navr\'atil}
\email{navratil@triumf.ca}
\affiliation{TRIUMF, 4004 Wesbrook Mall, Vancouver, British Columbia V6T 2A3, Canada}

\date{\today}

\begin{abstract}
Precision tests of the Standard Model and searches for beyond the Standard Model physics often require nuclear structure input. There has been a tremendous progress in the development of nuclear \textit{ab initio} techniques capable of providing accurate nuclear wave functions. For the calculation of observables, matrix elements of complicated operators need to be evaluated. Typically, these matrix elements would contain spurious contributions from the center-of-mass (c.m.) motion. This could be problematic when precision results are sought. Here, I derive a transformation relying on properties of harmonic oscillator wave functions that allows an exact removal of the c.m. motion contamination applicable to any one-body operator depending on nucleon coordinates and momenta. Resulting many-nucleon matrix elements are translationally invariant provided that the nuclear eigenfunctions factorize as products of the intrinsic and c.m. components as is the case, e.g., in the no-core shell model approach. An application of the transformation has been recently demonstrated in calculations of the nuclear structure recoil corrections for the $\beta$-decay of $^6$He.
\end{abstract} 
\maketitle

\section{Introduction}

Precision tests of the Standard Model (SM) and searches for beyond the Standard Model (BSM) physics that involve atomic nuclei typically require nuclear structure input. A prime example is the neutrinoless double $\beta$ decay~\cite{PhysRev.56.1184,PhysRevD.25.2951} where a knowledge of the nuclear transition matrix elements is needed to extract the neutrino mass if such a decay is observed. These matrix elements can be only obtained from nuclear theory~\cite{Engel_2017}. Recent years have brought advances in the sensitivity of $\beta$-decay studies. The $\beta$-decays are sensitive to interference of currents of SM particles and hypothetical BSM physics~\cite{CIRIGLIANO201393,GONZALEZALONSO2019165,cirigliano2019precision,hayen2020consistent,Falkowski2021}. The potential of using $\beta$-decay observables as the precision frontier for BSM searches has led to the deployment of several experimental efforts~\cite{doi:10.1063/1.4955362,PhysRevLett.114.162501,Ohayon2018,Mishnayot:2021uki}. However, discovering deviations from the SM predictions demands also high-precision theoretical calculations~\cite{Glick-Magid:2021uwb,Glick-Magid:2021xty,Sargsyan:2021jme}.

There has been a tremendous progress in the development of nuclear \textit{ab initio} techniques capable of providing accurate nuclear wave functions reaching masses of $A{=}100$~\cite{Gysbers2019NatPhys} and beyond. For the calculation of observables, matrix elements of complicated operators need to be evaluated. Typically, these matrix elements would contain spurious contributions from the center-of-mass (c.m.) motion. This could be problematic when precision results are sought. Motivated by ongoing and planned measurements of angular correlations between the emitted $\beta$-particles and the $\beta$-electron spectrum of $^6$He $\beta$ decay, {\it ab initio} no-core shell model (NCSM) calculations have been performed with the goal to determine as accurately as possible the Standard Model predictions for these observables~\cite{Glick-Magid:2021uwb}. As $^6$He is a light nucleus, a c.m. spurious contamination can be significant. Therefore, I developed a formalism to remove such contamination from matrix elements of arbitrary one-body operators including the complicated momentum transfer dependent operators relevant for the studied $\beta$-decay observables. It is the purpose of this paper to describe technical details of this formalism, which is a straightforward generalization of the calculation of the translationally invariant density of Ref.~\cite{PhysRevC.70.014317}. However, the present result is much more powerful with a broad applicability.

In Sect.~\ref{def}, coordinate and harmonic oscillator (HO) wave function transformations are introduced. The NCSM approach and the factorization of its eigenstates is discussed in Sect.~\ref{ncsm}. The derivation of the translationally invariant matrix elements of general one-body operators is presented in Sect.~\ref{deriv}. Applications with a particular focus on the $^6$He $\beta$-decay are reviewed in Sect.~\ref{appl} and conclusions are given in Sect.~\ref{concl}.

\section{Coordinate and HO wave function transformations}\label{def}

I follow the notation of Refs.~\cite{PhysRevC.70.014317} and~\cite{PhysRevC.61.044001}. For an $A$-nucleon system, one considers nucleons with the mass $m$ neglecting the difference between the proton and the neutron mass and use the following set of Jacobi coordinates: 
\begin{subequations}\label{jacobiam11}
\begin{eqnarray}
\vec{\xi}_0 &=& \textstyle{\sqrt{\frac{1}{A}}}\left[\vec{r}_1+\vec{r}_2
                                   +\ldots +\vec{r}_A\right]
\; , \\
\vec{\xi}_1 &=& \textstyle{\sqrt{\frac{1}{2}}}\left[\vec{r}_1-\vec{r}_2
                                                     \right]
\; , \\
\vec{\xi}_2 &=& \textstyle{\sqrt{\frac{2}{3}}}\left[\frac{1}{2}
                 \left(\vec{r}_1+\vec{r}_2\right)
                                   -\vec{r}_3\right]
\; , \\
&\ldots & \nonumber
\\
\vec{\xi}_{A-2} &=& \textstyle{\sqrt{\frac{A-2}{A-1}}}\left[\textstyle{\frac{1}{A-2}}
                    \left(\vec{r}_1+\vec{r}_2 + \ldots+ \vec{r}_{A-2}\right)  -\vec{r}_{A-1}\right]
\;\;\;\;\\
\vec{\xi}_{A-1} &=& \textstyle{\sqrt{\frac{A-1}{A}}}\left[\textstyle{\frac{1}{A-1}}
      \left(\vec{r}_1+\vec{r}_2 + \ldots+ \vec{r}_{A-1}\right)
                                   -\vec{r}_{A}\right] . 
\end{eqnarray}
\end{subequations}
Here, $\vec{\xi}_0$ is proportional to the center of mass of the $A$-nucleon system: $\vec{R}=\sqrt{\frac{1}{A}}\vec{\xi}_0$. 
On the other hand, $\vec{\xi}_\rho$ with $\rho{>}0$ is proportional to the relative position of the $\rho{+}1$-st nucleon and the
c.m. of the $\rho$ nucleons. Identical transformations can be introduced for the momenta $\vec{p}_i,\vec{\pi}_j$. The c.m. momentum $\vec{P}{=}\sum_{i=1}^A\vec{p}_i$ is then related to the $\vec{\pi}_0$ by $\vec{P}{=}\sqrt{A}\vec{\pi}_0$. In general, Jacobi coordinates are introduced as an orthogonal transformation of the single-nucleon coordinates and are proportional to differences of c.m. of nucleon sub-clusters. The particular choice (\ref{jacobiam11}) is convenient for the present derivations. For examples of other Jacobi coordinate sets see, e.g., Ref.~\cite{PhysRevC.61.044001}.

Let me rewrite the last and the first equation in (\ref{jacobiam11}) as
\begin{subequations}\label{jacobi_tr}
\begin{eqnarray}
\vec{\xi}_{A-1} &=& \textstyle{\sqrt{\frac{1}{A}}} \vec{R}^{A-1}_{\rm CM}
                    -\textstyle{\sqrt{\frac{A-1}{A}}}\vec{r}_{A}
\; , \\
\vec{\xi}_0 &=& \textstyle{\sqrt{\frac{A-1}{A}}} \vec{R}^{A-1}_{\rm CM}
                    +\textstyle{\sqrt{\frac{1}{A}}}\vec{r}_{A}     
\; ,
\end{eqnarray}
\end{subequations}
where 
$\vec{R}^{A-1}_{\rm CM}=\sqrt{\frac{1}{A-1}}\left[\vec{r}_1+\vec{r}_2+\ldots +\vec{r}_{A-1}\right]$.
Following, e.g., Ref. \cite{PhysRevC.5.1534}, the HO wave functions depending on the coordinates (\ref{jacobi_tr})
transform as
\begin{eqnarray}\label{ho_tr}
&&\sum_{M_1 m_1} (L_1 M_1 l_1 m_1|{\cal Q M_Q}) \varphi_{N_1 L_1 M_1}(\vec{R}^{A-1}_{\rm CM}) 
\varphi_{n_1 l_1 m_1}(\vec{r}_A) 
\nonumber \\
&=&\sum_{n l m N L M} \langle nl NL {\cal Q}|N_1 L_1 n_1 l_1 {\cal Q}\rangle_{\frac{1}{A-1}} \nonumber \\ 
&\times&    (l m L M|{\cal Q M_Q}) \, \varphi_{nlm}(\vec{\xi}_{A-1}) \varphi_{NLM}(\vec{\xi}_0)
\; ,
\end{eqnarray}
where $\langle nl NL {\cal Q}|N_1 L_1 n_1 l_1 {\cal Q}\rangle_{\frac{1}{A-1}}$ 
is the general HO bracket for two particles with mass ratio $\frac{1}{A-1}$ and ${\cal Q \; M_Q}$ the total angular momentum and its projection.

It should be noted that the same oscillator length, $b{=}b_0{=}\sqrt{\hbar/m\Omega}$, is used where $\Omega$ is the HO frequency, for all the HO wave functions appearing in Eq.~(\ref{ho_tr}) and throughout this paper, $\varphi_{nlm}(\vec{r}){=}R_{nl}(r,b_0)Y_{lm}(\hat{r})$ with $R_{nl}$ the radial HO wave function and $Y_{lm}$ the spherical harmonics and $\vec{r}$ any of the coordinates in Eqs.~(\ref{jacobiam11}) and (\ref{jacobi_tr}).

\section{Factorization of no-core shell model eigenstates}\label{ncsm}

Expansions on square integrable many-body basis states are among the most common techniques for the description of the static properties of atomic nuclei. The HO basis is frequently utilized.

The {\it ab initio} NCSM~\cite{PhysRevLett.84.5728,PhysRevC.62.054311,Navratil2009,Barrett2013} is one of such techniques. Nuclei are considered as systems of $A$ non-relativistic point-like nucleons interacting through realistic inter-nucleon interactions typically derived using the chiral effective filed theory (EFT)~\cite{Weinberg1990}. All nucleons are active degrees of freedom. Translational invariance as well as angular momentum and parity of the system under consideration are conserved. The many-body wave function is cast into an expansion over a complete set of antisymmetric $A$-nucleon HO basis states containing up to  $N_{\rm max}$ HO excitations above the lowest possible configuration. The basis is further characterized by the frequency $\Omega$ of the HO well and may depend on either Jacobi~\cite{PhysRevC.61.044001} or single-particle coordinates~\cite{PhysRevC.62.054311}. In the former case, the wave function does not contain the c.m. motion, but antisymmetrization is complicated. In the latter case, antisymmetrization is trivially achieved using Slater determinants (SD), but the c.m. degrees of freedom are included in the basis. Calculations with the two alternative coordinate choices are completely equivalent.

Square-integrable energy eigenstates are obtained by solving the Schr\"{o}dinger equation
\begin{equation}\label{NCSM_eq}
\hat{H} \ket{A \lambda J^\pi M T T_z} = E_{\lambda\,T_z}^{J^\pi T} \ket{A \lambda J^\pi M T T_z},
\end{equation}
with the intrinsic Hamiltonian
\begin{equation}\label{eq:ham_Ham}
\hat{H}=\frac{1}{A}\sum_{i<j=1}^A\frac{(\vec{p}_i-\vec{p}_j)^2}{2m}
+ \sum_{i<j=1}^A \hat{V}^{NN}_{ij} + \sum_{i<j<k=1}^A
\hat{V}^{3N}_{ijk}.
\end{equation}
Here, $\vec{p}$ are nucleon momenta, $\hat{V}^{NN}$ is the nucleon-nucleon (\(N\!N\)) and $\hat{V}^{3N}$ the three-nucleon (3N) interaction. The $\lambda$ in~\eqref{NCSM_eq} labels eigenstates with identical $J^\pi T T_z$. In general, the isospin $T$ is only approximately conserved.

Calculations in the SD basis are typically more efficient for nuclei with $A{>}4$. The so-called $M$-scheme is then often used with the basis characterized by the angular momentum projection $M$, the parity $\pi$, and $T_z{=}(Z-N)/2$. The eigenstates are then obtained by applying the Lanczos algorithm~\cite{Lanczos1950}. The relationship between the Jacobi coordinate and the SD eigenstates is 
\begin{eqnarray}\label{state_relation}
&&\langle \vec{r}_1 \ldots \vec{r}_A \sigma_1 \ldots \sigma_A \tau_1 \ldots \tau_A 
  | A \lambda J^\pi M T T_z\rangle_{\rm SD} \\ \nonumber
  &=& 
\langle \vec{\xi}_1 \ldots \vec{\xi}_{A-1} \sigma_1 \ldots \sigma_A \tau_1 \ldots \tau_A 
| A \lambda J^\pi M T T_z\rangle \varphi_{000}(\vec{\xi}_0) 
\end{eqnarray}
with the $\sigma$ and $\tau$ the spin and isospin coordinates, respectively. To simplify the notation, I will omit the isospin coordinates and quantum numbers and the parity from now on. In order to select the physical SD eigenstates satisfying the factorization (\ref{state_relation}), one typically applies the Lawson projection~\cite{GLOECKNER1974313} that pushes SD eigenstates with the c.m. in excited HO configurations to high energy relative to the c.m. $0\hbar\Omega$ eigenstates. 

It should be noted that the factorization (\ref{state_relation}) is not unique to the NCSM approach, an analogous relation was found, e.g., in converged coupled cluster calculations~\cite{PhysRevLett.103.062503}.  

\section{Translationally invariant matrix elements}\label{deriv}

Let me consider a general one-body operator
\begin{equation}\label{opKunph}
\widehat{O}^{\left( K\right)} = \sum_i^A \widehat{O}^{\left(K\right)}(\vec{r}_i,\sigma_i) \; ,
\end{equation}
that might also depend on nucleon momenta $\vec{p}_i$. A physical nuclear operator is expected to be translationally invariant, i.e., the coordinates should be measured from the c.m. of the nucleus rather than from the origin of the HO well (or other potential used to define the single-particle basis). Consequently, one should rather consider the operator~\cite{PhysRev.108.482,Greiner1988,Baye_2018,PhysRevC.61.044001}
\begin{equation}\label{opK}
\widehat{O}^{\left( K\right)} = \sum_i^A \widehat{O}^{\left(K\right)}(\vec{r}_i-\vec{R},\sigma_i) 
\end{equation}
with $\vec{R}$ the c.m. coordinate of the nucleus. I note that the operator~(\ref{opK}) may also depend on nucleon momenta, $\vec{p}_i-\textstyle{\frac{1}{A}}\vec{P}$, with $\vec{P}$ the c.m. momentum of the nucleus. To simplify the notation, I do not show this dependence explicitly.

As the nuclear wave functions are antisymmetric with respect to exchanges of the nucleons, a matrix element of the operator (\ref{opK}) can be evaluated as~\cite{PhysRevC.61.044001}
\begin{eqnarray}\label{physME}
  &&\langle A\lambda_{f} J_{f}|| \widehat{O}^{\left( K\right)} || A\lambda_{i}J_{i}\rangle   \\ \nonumber
  &=& A\; \langle A\lambda_{f} J_{f}|| \widehat{O}^{\left( K\right)}(\vec{r}_A-\vec{R},\sigma_A) || A\lambda_{i}J_{i}\rangle   \\ \nonumber
  &=& A\; \langle A\lambda_{f} J_{f}|| \widehat{O}^{\left( K\right)}\left(-\textstyle{\sqrt{\frac{A-1}{A}}}\vec{\xi},\sigma \right) || A\lambda_{i}J_{i}\rangle \; ,
\end{eqnarray}
where $\langle||\hat{O}||\rangle$ denotes a reduced matrix element in angular momentum, $\vec{\xi} {=} \vec{\xi}_{A-1}$ with $\vec{\xi}_{A-1} $ given by Eq.~(\ref{jacobiam11}e) and $\sigma {=} \sigma_A$. If the operator depends explicitly on nucleon momenta $\vec{p}_i-\textstyle{\frac{1}{A}}\vec{P}$, a Jacobi momentum $\vec{\pi}{=} \vec{\pi}_{A-1}$ will appear in Eq.~(\ref{physME}) with the same scaling factor as that of $\vec{\xi}$, i.e., $-\textstyle{\sqrt{\frac{A-1}{A}}}\vec{\pi}$.

Let me re-write the matrix element (\ref{physME}) as integral over the Jacobi coordinates and introduce a single-particle-like matrix element of the operator:
\begin{widetext}
\begin{eqnarray}\label{physMEinteg}
  &&\langle A\lambda_{f} J_{f}|| \widehat{O}^{\left( K\right)} || A\lambda_{i}J_{i}\rangle   
  = A \textstyle{\frac{1}{\widehat{K}}} \hat{J}_f \sum (J_i M_i K k | J_f M_f) 
     \langle n l j || \widehat{O}^{\left( K\right)}\left(-\textstyle{\sqrt{\frac{A-1}{A}}}\vec{\xi},\sigma \right) || n' l' j' \rangle
  \\ \nonumber
  &\times&  (-1)^{j'-m'_{j}} (j m j' -m'_j | K k) 
  (l m \textstyle{\frac{1}{2}} m_s | j m_j) (l' m' \textstyle{\frac{1}{2}} m'_{s} | j' m'_j) 
  \int d\vec{\xi}_1\ldots d\vec{\xi}_{A-2} d\vec{\xi}_{A-1} d\vec{\xi}'_{A-1} \\ \nonumber
   &\times&  \langle A\lambda_{f} J_{f} M_f | \vec{\xi}_1\ldots \vec{\xi}_{A-2}\vec{\xi}_{A-1}\sigma_1 \ldots\sigma_{A-1}\sigma_A \rangle  
    \varphi_{nlm}(\vec{\xi}_{A-1}) \chi_{m_s}(\sigma_A) \varphi^*_{n'l'm'}(\vec{\xi}'_{A-1}) \chi^*_{m'_s}(\sigma'_A) \\ \nonumber
  &\times& \langle  \vec{\xi}_1\ldots \vec{\xi}_{A-2}\vec{\xi}'_{A-1}\sigma_1 \ldots\sigma_{A-1}\sigma'_A | A\lambda_{i}J_{i} M_i \rangle \; .
\end{eqnarray}
The sum runs over all quantum numbers that do not appear on the left hand side except $M_f$. An abbreviation $\hat{J}{=}\sqrt{2J+1}$ is used. It should be noted that the ket and bra states in the single-particle-like matrix element $ \langle n l j || \widehat{O}^{\left( K\right)}\left(-\textstyle{\sqrt{\frac{A-1}{A}}}\vec{\xi},\sigma \right) || n' l' j' \rangle$ depend on the Jacobi coordinate $\vec{\xi}$ rather than on a regular coordinate $\vec{r}$ like in a standard one-body operator matrix element. In deriving Eq. (\ref{physMEinteg}) I used the Dirac $\delta$ function properties and the completeness relations
$\delta(\vec{\xi}-\vec{\xi}_{A-1}){=}\sum_{nlm} \varphi_{nlm}(\vec{\xi}_{A-1}) \varphi_{nlm}^*(\vec{\xi})$ and $\delta_{\sigma\sigma_A}{=}\sum_{m_s} \chi_{m_s}(\sigma_A)\chi^*_{m_s}(\sigma)$.

I aim at calculating the matrix element (\ref{physMEinteg}) with the help of the SD eigenstates (\ref{state_relation}) that are obtained more efficiently for nuclei with $A{>}4$ usually by applying the second quantization. To achieve that goal, I investigate an analogous integral to that appearing on the right-hand side of Eq.~(\ref{physMEinteg}) for the Cartesian coordinate HO wave functions (summing over quantum numbers that do not appear on the right hand side),
\begin{eqnarray}\label{SDrel1}
  &&\sum (-1)^{j_2-m_2} (j_1 m_1 j_2 -m_2 | K k) 
  \int d\vec{r}_1\ldots d\vec{r}_A d\vec{r}'_{A} \,  _{\rm SD}\langle A\lambda_{f} J_{f} M_f | \vec{r}_1\sigma_1\ldots \vec{r}_{A}\sigma_{A}\rangle \\ \nonumber
  &\times& \varphi_{n_1 l_1 j_1 m_1}(\vec{r}_A\sigma_A)  \varphi^*_{n_2 l_2 j_2 m_2}(\vec{r}'_A\sigma'_A) 
   \langle \vec{r}_1\sigma_1\ldots \vec{r}'_{A}\sigma'_{A}| A\lambda_{i}J_{i} M_i \rangle_{\rm SD} \\ \nonumber
  &=& - \textstyle{\frac{1}{A}} \textstyle{\frac{1}{\hat{J}_f}} (J_i M_i K k | J_f M_f) 
   _{\rm SD}\langle A\lambda_{f} J_{f}||(a^\dagger_{n_1 l_1 j_1} \tilde{a}_{n_2 l_2 j_2})^{(K)}||A\lambda_{i}J_{i} \rangle_{\rm SD} \; ,
\end{eqnarray}
with $\varphi_{n l j m}(\vec{r}\sigma) {=}\sum_{m_l m_s}(l m_l \textstyle{\frac{1}{2}} m_s | j m) \varphi_{n l m_l}(\vec{r}) \chi_{m_s}(\sigma)$, and $a^\dagger_{n l j m}$, $\tilde{a}_{n l j m}{=}(-1)^{j-m} a_{n l j, -m}$ the creation and annihilation operators. The second quantization matrix elements, $(-1/\widehat{K}) \; _{\rm SD}\langle A\lambda_{f} J_{f}||(a^\dagger_{n_1 l_1 j_1} \tilde{a}_{n_2 l_2 j_2})^{(K)}||A\lambda_{i}J_{i} \rangle_{\rm SD}$, contain the many-body structure information and are typically referred to as the one-body density matrix elements (OBDME).

Next, I re-write the left-hand-side of Eq.~(\ref{SDrel1}) and perform a change of variables to the Jacobi coordinates using Eqs.~(\ref{jacobiam11})-(\ref{ho_tr}), (\ref{state_relation}), together with 
$\delta(\vec{R}^{A-1}_{\rm CM}-\vec{R}^{\prime A-1}_{\rm CM})=
\sum_{N_1 L_1 M_1} \varphi_{N_1 L_1 M_1}(\vec{R}^{A-1}_{\rm CM}) 
\varphi_{N_1 L_1 M_1}^*(\vec{R}^{\prime A-1}_{\rm CM})$ (see also Eq.~(12) in Ref.~\cite{PhysRevC.70.014317}),
\begin{eqnarray}\label{SDrel2}
  &&\sum (-1)^{j_2-m_2} (j_1 m_1 j_2 -m_2 | K k) 
 \int d\vec{r}_1\ldots d\vec{r}_A d\vec{r}'_{A} \,  _{\rm SD}\langle A\lambda_{f} J_{f} M_f |\vec{r}_1\sigma_1\ldots \vec{r}_{A}\sigma_{A}\rangle \\ \nonumber
  &\times&  \varphi_{n_1 l_1 j_1 m_1}(\vec{r}_A\sigma_A)  \varphi^*_{n_2 l_2 j_2 m_2}(\vec{r}'_A\sigma'_A) 
  \langle \vec{r}_1\sigma_1\ldots \vec{r}'_{A}\sigma'_{A}| A\lambda_{i}J_{i} M_i \rangle_{\rm SD} \\ \nonumber
  &=& \sum \hat{j}_1 \hat{j}_2 \hat{j} \hat{j}' \hat{l} \hat{l}' (-1)^{K+L_1+l_1+l_2+j'+j_2} 
\left\{ \begin{array}{ccc} j' & L_1 & j_2 \\
  l_2  & \textstyle{\frac{1}{2}} & l'   
\end{array}\right\}
\left\{ \begin{array}{ccc} j_1 & L_1 & j \\
  l  & \textstyle{\frac{1}{2}} & l_1   
\end{array}\right\}                                          
\left\{ \begin{array}{ccc} j_1 & L_1 & j \\
  j'  & K & j_2   
\end{array}\right\} \\ \nonumber
&\times&\langle n l 0 0 l | N_1 L_1 n_1 l_1 l\rangle_{\frac{1}{A-1}}
 \langle n' l' 0 0 l' | N_1 L_1 n_2 l_2 l'\rangle_{\frac{1}{A-1}} 
   (-1)^{j'-m'_j} (j m j' -m'_j | K k) \\ \nonumber
  &\times& (l m \textstyle{\frac{1}{2}} m_s | j m_j) (l' m' \textstyle{\frac{1}{2}} m'_{s} | j' m'_j) 
  \int d\vec{\xi}_1\ldots d\vec{\xi}_{A-1} d\vec{\xi}'_{A-1}  
           \langle A\lambda_{f} J_{f} M_f | \vec{\xi}_1\ldots \vec{\xi}_{A-1}\sigma_1 \ldots\sigma_A \rangle  \\ \nonumber
  &\times&   \varphi_{nlm}(\vec{\xi}_{A-1}) \chi_{m_s}(\sigma_A) \varphi^*_{n'l'm'}(\vec{\xi}'_{A-1}) \chi^*_{m'_s}(\sigma'_A) 
  \langle  \vec{\xi}_1\ldots \vec{\xi}'_{A-1}\sigma_1 \ldots\sigma'_A| A\lambda_{i}J_{i} M_i\rangle  \\ \nonumber
  &=& \sum  \left( M^{K}\right)_{n_{1}l_{1}j_{1}n_{2}l_{2}j_{2},nlj n'l'j'} 
     (-1)^{j'-m'_{j'}} (j m j' -m'_j | K k) 
  (l m \textstyle{\frac{1}{2}} m_s | j m_j) (l' m' \textstyle{\frac{1}{2}} m'_{s} | j' m'_j) 
  \int d\vec{\xi}_1\ldots d\vec{\xi}_{A-1} d\vec{\xi}'_{A-1} \\ \nonumber
   &\times& \langle A\lambda_{f} J_{f} M_f | \vec{\xi}_1\ldots \vec{\xi}_{A-1}\sigma_1 \ldots\sigma_A \rangle  
  \varphi_{nlm}(\vec{\xi}_{A-1}) \chi_{m_s}(\sigma_A) \varphi^*_{n'l'm'}(\vec{\xi}'_{A-1}) \chi^*_{m'_s}(\sigma'_A) 
   \langle  \vec{\xi}_1\ldots \vec{\xi}'_{A-1}\sigma_1 \ldots\sigma'_A| A\lambda_{i}J_{i} M_i \rangle \; ,
\end{eqnarray}
\end{widetext}
where I introduced the matrix
\begin{eqnarray}\label{M_K}
  &&(M^K)_{n_1 l_1 j_1 n_2 l_2 j_2, n l j n' l' j'}   \\ \nonumber
  &=& \sum_{N_1 L_1} \hat{j}_1 \hat{j}_2 \hat{j} \hat{j}' \hat{l} \hat{l}' (-1)^{K+L_1+l_1+l_2+j'+j_2} \\ \nonumber
&\times&
\left\{ \begin{array}{ccc} j' & L_1 & j_2 \\
  l_2  & \textstyle{\frac{1}{2}} & l'   
\end{array}\right\}
\left\{ \begin{array}{ccc} j_1 & L_1 & j \\
  l  & \textstyle{\frac{1}{2}} & l_1   
\end{array}\right\}                                          
\left\{ \begin{array}{ccc} j_1 & L_1 & j \\
  j'  & K & j_2   
\end{array}\right\} \\ \nonumber
&\times&\langle n l 0 0 l | N_1 L_1 n_1 l_1 l\rangle_{\frac{1}{A-1}}
\langle n' l' 0 0 l' | N_1 L_1 n_2 l_2 l'\rangle_{\frac{1}{A-1}}
\; .
\end{eqnarray}
The dimension of the matrix is given by all the combinations of two-nucleon HO states $\{nlj\}_1 \{nlj\}_2$ with the $j_1, j_2$ coupled to $K$ satisfying truncation criteria for the $N_{\rm max}$ model space. The transformation matrix $M^K$ (\ref{M_K}) is a straightforward generalization of the matrix introduced in Eq.~(13) of Ref.~\cite{PhysRevC.70.014317}.

Combining Eqs.~(\ref{physMEinteg})-(\ref{SDrel2}), one arrives at the final result
\begin{eqnarray}\label{trinvME}
  &&\langle A\lambda_{f} J_{f}|| \widehat{O}^{\left( K\right)} || A\lambda_{i}J_{i}\rangle \\ \nonumber
  &=&-\textstyle{\frac{1}{\widehat{K}}}\sum \langle nlj\left\| \widehat{O}^{\left( K\right) }\left( -\textstyle{\sqrt{\frac{A-1}{A}}}\vec{\xi },\sigma \right) \right\| n'l'j'\rangle \\ \nonumber
  &\times&\left( M^{K}\right)^{-1} _{nlj n'l'j',n_{1}l_{1}j_{1}n_{2}l_{2}j_{2}} \\ \nonumber
  &\times& _{\rm SD}\langle A\lambda_{f} J_{f}|| (a^\dagger_{n_1l_1j_1} \tilde{a}_{n_2l_2j_2}) ^{(K)} || A\lambda_{i}J_{i}\rangle_{\rm SD} \; .
\end{eqnarray}
This relation can be contrasted with the ``standard'' many-body matrix element of a one-body operator obtained in the SD basis with the c.m. motion included
\begin{eqnarray}\label{SDME}
  &&_{\rm SD}\langle A\lambda_{f} J_{f}|| \widehat{O}^{\left( K\right)} || A\lambda_{i}J_{i}\rangle_{\rm SD}  \\ \nonumber
  &=&-\textstyle{\frac{1}{\widehat{K}}}\sum \langle n_1 l_1 j_1 \left\| \widehat{O}^{\left( K\right) }\left(\vec{r},\sigma \right) \right\| n_2 l_2 j_2\rangle \\ \nonumber
  &\times& _{\rm SD}\langle A\lambda_{f} J_{f}|| (a^\dagger_{n_1l_1j_1} \tilde{a}_{n_2l_2j_2})^{(K)} || A\lambda_{i}J_{i}\rangle_{\rm SD} \; .
\end{eqnarray}

In Eq.~(\ref{trinvME}), the OBDME contain the many-body nuclear structure information, the transformation matrix $M^K$ removes the spurious c.m. contributions from the OBDME, and the operator action is given in the single-particle-like matrix elements depending on the Jacobi coordinates $\vec{\xi}$.  

\section{Applications}\label{appl}

\subsection{$^6$He$\rightarrow ^6$Li $\beta$ decay}
  
The formalism developed in this paper was applied for the first time in calculations of the electron spectrum of the $^6$He$\rightarrow ^6$Li $\beta$ decay~\cite{Glick-Magid:2021uwb}. In general, nuclear electroweak processes can be described in terms of seven basic multipole operators depending on the transferred momentum~\cite{Donnelly:1979ezn}. Four of these operators are relevant for the  $^6$He $\beta$ decay calculations,
\begin{equation}
  \label{eq:operators}
  \begin{split}
    \hat{\Sigma}^{\prime\prime}_{JM_J}(q \vec{r}_j) &= \left[
      \frac{1}{q} \vec{\nabla}_{\vec{r}_j} M_{JM_J}(q
      \vec{r}_j)  \right] \cdot \vec{\sigma}_j,\\
    \hat{\Omega}^\prime_{JM_J}(q \vec{r}_j) &= M_{JM_J}(q \vec{r}_j)
    \,\vec{\sigma}_j \cdot \vec{\nabla}_{\vec{r}_j} +
    \frac{1}{2}\hat{\Sigma}^{\prime\prime}_{JM_J}(q \vec{r}_j),\\
    \hat{\Delta}_{JM_J}(q \vec{r}_j) &= \vec{M}_{JJM_J}(q \vec{r}_j)
    \cdot \frac{1}{q}\vec{\nabla}_{\vec{r}_j}, \\
    \hat{\Sigma}^{\prime}_{JM_J}(q \vec{r}_j) &= -i \left[ \frac{1}{q}
      \vec{\nabla}_{\vec{r}_j} \times \vec{M}_{JJM_J}(q
      \vec{r}_j) \right] \cdot \vec{\sigma}_j,
  \end{split}
\end{equation}
with \(\vec{\sigma}_j\) being the Pauli spin matrices associated with nucleon \(j\) and $q$ the magnitude of the transfer momentum. Furthermore, \(M_{JM_J}(q \vec{r}_j) = j_J(q r_j) Y_{JM_J}({\hat{r}_j})\) and \(\vec{M}_{JLM_J}(q \vec{r}_j) = j_L(q r_j) \vec{Y}_{JLM_J}({\hat{r}_j})\), where \({\hat{r}_j}\) represents azimuthal and polar angles of \(\vec{r}_j\). \(Y_{JM_J}\) and \(\vec{Y}_{JLM_J}\) are spherical and vector spherical harmonics, and \(j_J\) are spherical Bessel functions.

Results for the nuclear matrix elements of the one-body operators~\eqref{eq:operators} are shown in the figure in the Supplemental Material of Ref.~\cite{Glick-Magid:2021uwb}. These matrix elements are then used to construct the nuclear structure input for the electron spectrum calculations, matrix elements of the longitudinal axial current $\hat{L}^A $, axial charge $\hat{C}^A$, and vector magnetic $\hat{M}^V$ operators, given in Ref.~\cite{Glick-Magid:2021uwb}. One observes that the translational-invariant many-body matrix elements, Eq.~(\ref{trinvME}), and the standard many-body matrix elements, Eq.~(\ref{SDME}), are the same at $q{=}0$ for the $\hat{\Sigma}^\prime$, $\hat{\Sigma}^{\prime\prime}$, and $\hat{\Delta}$ operators while the many-body matrix elements of the $\hat{\Omega}^\prime$ differ by about a factor of two. In particular, the spurious center-of-mass components of the wave functions contaminate the matrix elements when the gradient in the first term of the $\hat{\Omega}^\prime$ is applied on the wave function. The overall effect is quite significant. With an increasing $q$, all the operators become contaminated by spurious c.m. contributions although the effect is small for $\hat{\Sigma}^\prime$, $\hat{\Sigma}^{\prime\prime}$, and $\hat{\Delta}$ in the $q$ range relevant for the $^6$He $\beta$ decay.

When evaluating the one-body-like Jacobi-coordinate matrix elements appearing in Eq.~(\ref{trinvME}) of the operators (\ref{eq:operators}), one first carries out the gradients in the parenthesis of the $\hat{\Sigma}^\prime$, $\hat{\Sigma}^{\prime\prime}$ (see, e.g., Refs.~\cite{Donnelly:1979ezn,HAXTON2008345}) and then replaces $\vec{r}$ by $ -\textstyle{\sqrt{\frac{A-1}{A}}}\vec{\xi }$ in all the operators, and, in addition, one replaces the gradients in $\hat{\Omega}^\prime$ and $\hat{\Delta}$ by $ -\textstyle{\sqrt{\frac{A-1}{A}}}\vec{\nabla}_{\vec{\xi }}$. The latter step corresponds to the replacement of the momentum $\vec{p}$  by  $ -\textstyle{\sqrt{\frac{A-1}{A}}}\vec{\pi }$.

\subsection{Matrix elements of the seven basic multipole operators}\label{Sevenop}

It should be noted that one-body matrix elements of the seven basic multipole operators for electroweak processes can be carried out analytically in the HO basis as demonstrated in Refs.~\cite{Donnelly:1979ezn,HAXTON2008345}. In Ref.~\cite{HAXTON2008345}, a Mathematica script is provided for the calculation of the matrix elements. These results can be readily applied to calculate the matrix elements of the translationally invariant versions of the operators used here. In the analytic results, e.g., Eqs.~(17)-(19) in Ref.~\cite{HAXTON2008345}, (i) the $q$ is replaced by $-\textstyle{\sqrt{\frac{A-1}{A}}} q$, (ii) the matrix elements (18) and (19) in Ref.~\cite{HAXTON2008345} are multiplied by one more factor of $-\textstyle{\sqrt{\frac{A-1}{A}}}$ due to the gradient (momentum) acting on the wave function, and, finally, (iii) yet another factor of $-\textstyle{\sqrt{\frac{A-1}{A}}}$ is applied to terms with $1/q$, i.e., $\hat{\Sigma}^\prime$, $\hat{\Sigma}^{\prime\prime}$, and $\hat{\Delta}$ (\ref{eq:operators}), to compensate for the extra scaling in step (i).

\subsection{Electromagnetic transitions}\label{EMop}

Electromagnetic transitions between nuclear states can typically be described in long wave length approximation using simple multipole operators, e.g., $EL\approx e_p r_p^L Y_L(\hat{r}_p)+e_n r_n^L Y_L(\hat{r}_n)$ with $L{=}1$ for electric dipole, $L{=}2$ for electric quadrupole, etc., and $M1\approx g_{l_p} \vec{l}_p + g_{l_n} \vec{l}_n + g_{s_p} \vec{s}_p + g_{s_n} \vec{s}_n$. For these operators, applications of Eqs.~(\ref{trinvME}) and (\ref{SDME}) lead to identical results as long as $L{>}0$. When calculating the one-body-like matrix elements in Eq.~(\ref{trinvME}), one substitutes $r^L Y_L(\hat{r})\rightarrow (-\textstyle{\sqrt{\frac{A-1}{A}}})^L \xi^L Y_L(\hat{\xi})$ and $\vec{l}=\vec{r}\times\vec{p}\rightarrow \textstyle{\frac{A-1}{A}} \vec{\xi}\times\vec{\pi}$. There is obviously no scaling for the spin operators.

\subsection{Nuclear radii and kinetic energy}\label{Radop}

Point proton, neutron, and matter radii are obtained by calculating mean values of the operators $\textstyle{\frac{1}{Z}}\sum_{i=1}^A (\vec{r}_i-\vec{R})^2 (1/2+t_{z,i})$, $\textstyle{\frac{1}{N}}\sum_{i=1}^A (\vec{r}_i-\vec{R})^2 (1/2-t_{z,i})$, and $\textstyle{\frac{1}{A}}\sum_{i=1}^A (\vec{r}_i-\vec{R})^2$, respectively. These are not one-body operators and to calculate their matrix elements, one typically re-writes them in a two-body-like form proportional to $(\vec{r}_i-\vec{r}_j)^2$, see, e.g., Ref.~\cite{Caprio_2020}. The many-body matrix elements are then obtained with the help of two-body density matrix elements, i.e., $\langle a^\dagger a^\dagger a a \rangle$. The present formalism allows a much simpler evaluation of the radii using only the OBDME. The one-body-like matrix element in Eq.~(\ref{trinvME}) is obtained in an analytical form using $(\vec{r}-\vec{R})^2\rightarrow \textstyle{\frac{A-1}{A}} \xi^2$:
\begin{eqnarray}\label{eqrad}
  \langle nlj || \xi^2 || nlj\rangle &=& \hat{j}\, (2 n + l + 3/2)\, b^2 \\ \nonumber
  \langle nlj || \xi^2 || n{+}1lj\rangle &=& -\hat{j}\, \sqrt{(n+1)(n+l+3/2)}\, b^2 \\ \nonumber
  &=& \langle n{+}1lj || \xi^2 || nlj\rangle \; , 
\end{eqnarray}
and zero otherwise. Here, $b^2{=}\frac{\hbar}{m\Omega}$ and the $\hat{j}{=}\sqrt{2j+1}$ is due to the fact that the matrix element is reduced in $j$. The point-proton (neutron, matter) radius is obtained using the proton (neutron, proton plus neutron) OBDME, i.e., $\langle a^\dagger_p a_p\rangle$ ($\langle a^\dagger_n a_n\rangle$, $\langle a^\dagger_p a_p\rangle{+}\langle a^\dagger_n a_n\rangle$) with $1/Z$($1/N$, $1/A$) times $\textstyle{\frac{A-1}{A}}$ scaling.

The kinetic energy can be calculated in an analogous way, starting from the operator $\textstyle{\frac{1}{2m}}\sum_{i=1}^A (\vec{p}_i-\textstyle{\frac{1}{A}}\vec{P})^2$, replacing $(\vec{p}-\textstyle{\frac{1}{A}}\vec{P})^2\rightarrow \textstyle{\frac{A-1}{A}} \pi^2$. The one-body-like HO matrix elements of $\textstyle{\frac{1}{2m}}\pi^2$ are obtained as in Eq.~(\ref{eqrad}) with $b^2$ replaced by $\hbar\Omega/2$ and a positive sign in the off-diagonal term. In this case, one sums the proton and the neutron OBDME and scales by $\textstyle{\frac{A-1}{A}}$. Again, a standard kinetic energy calculation would require a two-body density. The present calculation is much simpler.

\subsection{Nuclear density}\label{density}

To obtain the nuclear density, one uses the operator $\rho(\vec{r}-\vec{R}){=}\sum_{i=1}^A \delta(\vec{r}-\vec{R}-(\vec{r}_i-\vec{R}))$ and substitutes $\vec{r}-\vec{R}\rightarrow -\textstyle{\sqrt{\frac{A-1}{A}}}\vec{\xi}$ as done in Ref.~\cite{PhysRevC.70.014317}. Using the density operator, i.e., $\rho(\vec{\xi}){=}\delta(\vec{\xi}-\vec{\xi}_{A-1})$ in Eq.~(\ref{trinvME}) with the HO states $\langle \vec{\xi}_{A-1}\sigma|nlj\rangle$, results numerically in the same local translationally invariant density as that obtained from Eq.~(16) of Ref.~\cite{PhysRevC.70.014317}. This is not obvious as the $\rho(\vec{\xi})$ expression from Eq.~(\ref{trinvME}) does not simply reduce to Eq.~(16) of  Ref.~\cite{PhysRevC.70.014317}.

The non-local translationally invariant density is obtained using the operator $\delta(\vec{\xi}-\vec{\xi}_{A-1})\;\delta(\vec{\xi}'-\vec{\xi}'_{A-1})$ in Eq.~(\ref{trinvME}). Numerically, the results are identical to those obtained from Eq.~(16) of Ref.~\cite{PhysRevC.97.034619}. Again, this is not obvious as  Eq.~(\ref{trinvME}) does not simply reduce to  Eq.~(16) of Ref.~\cite{PhysRevC.97.034619}. From the non-local density one can derive the translationally invariant kinetic density~\cite{PhysRevC.99.024305}, which the present formalism should also simplify.

The present formalism can be readily applied to calculate spin-dependent density, see, e.g., Ref.~\cite{PhysRevC.103.054314}, to construct optical potentials for high-energy nucleon scattering on nuclei. Calculations in this direction are under way.

\section{Conclusions}\label{concl}

I derived an expression for calculations of translationally invariant nuclear matrix elements for arbitrary one-body operators. The derivation relies on the factorization of the c.m. and intrinsic components of the nuclear eigenstates and utilizes properties of HO wave functions. The main result given in Eq.~(\ref{trinvME}) is a straightforward generalization of the formalism derived in Ref.~\cite{PhysRevC.70.014317} for nuclear density. It is, however, much more powerful with a much wider applicability. It has been already successfully applied to calculations of nuclear structure recoil corrections to $^6$He $\beta$ decay in Ref.~\cite{Glick-Magid:2021uwb} within the NCSM.

The present formalism is in particular relevant for momentum transfer dependent operators such as the electroweak seven basic multipole operators ~\cite{Donnelly:1979ezn} or operators relevant for the hypothetical dark matter scattering off nuclei~\cite{Fitzpatrick:2012ix,Gazda:2016mrp} as the c.m. contamination of matrix elements typically increases with the transferred momentum. At the same time, Eq.~(\ref{trinvME}) in combination with Eq.~(\ref{eqrad}) allows the calculation of nuclear radii as well as the kinetic energy using only the OBDME contrary to a traditional approach where two-body density matrix elements are required.

In this paper, I discussed applications within the NCSM method. The c.m. and intrinsic wave function factorization is achieved also in other {\it ab initio} techniques, e.g., the coupled cluster method (CCM)~\cite{PhysRevLett.103.062503,Hagen_2016} or in-medium similarity renormalization group (IMSRG) method~\cite{PhysRevLett.106.222502,doi:10.1146/annurev-nucl-101917-021120}, at least to a good approximation. It remains to be seen how successfully the present formalism can be applied in such techniques.

\acknowledgments

I thank Christian Forss\'en, Daniel Gazda, Michael Gennari, Peter Gysbers, and J\'er\'emy Dohet-Eraly for useful discussions. This work was supported by the NSERC Grant No. SAPIN-2016-00033. TRIUMF receives federal funding via a contribution agreement with the National Research Council of Canada.  Computing support came from an INCITE Award on the Summit supercomputer of the Oak Ridge Leadership Computing Facility (OLCF) at ORNL, and from Westgrid and Compute Canada.

\end{document}